\documentclass[twocolumn,british,pra, reprint]{revtex4}
\usepackage[T1]{fontenc}
\usepackage[latin9]{inputenc}
\usepackage[letterpaper]{geometry}
\usepackage{amssymb}
\usepackage{graphicx}
\usepackage{esint}
\usepackage{hyperref}

\makeatletter
\@ifundefined{textcolor}{}
{%
 \definecolor{BLACK}{gray}{0}
 \definecolor{WHITE}{gray}{1}
 \definecolor{RED}{rgb}{1,0,0}
 \definecolor{GREEN}{rgb}{0,1,0}
 \definecolor{BLUE}{rgb}{0,0,1}
 \definecolor{CYAN}{cmyk}{1,0,0,0}
 \definecolor{MAGENTA}{cmyk}{0,1,0,0}
 \definecolor{YELLOW}{cmyk}{0,0,1,0}
}

\makeatother

\usepackage{babel}
\begin{document}

\pagebreak{}

\title{Time Delay in Molecular Photoionization}

\author{P. Hockett}

\email{paul.hockett@nrc.ca}

\affiliation{National Research Council of Canada, 100 Sussex Drive, Ottawa, K1A
0R6, Canada}

\author{E. Frumker}

\affiliation{Department of Physics, Ben-Gurion University of the Negev, Beer-Sheva
84105, Israel}

\author{D.M. Villeneuve, P.B. Corkum}

\affiliation{Joint Attosecond Science Laboratory, National Research Council of
Canada and University of Ottawa, 100 Sussex Drive, Ottawa, K1A 0R6,
Canada}
\begin{abstract}
Time-delays in the photoionization of molecules are investigated.
As compared to atomic ionization, the time-delays expected from molecular
ionization present a much richer phenomenon, with a strong spatial
dependence due to the anisotropic nature of the molecular scattering
potential. We investigate this from a scattering theory perspective,
and make use of molecular photoionization calculations to examine
this effect in representative homonuclear and hetronuclear diatomic
molecules, nitrogen and carbon monoxide. We present energy and angle-resolved
maps of the Wigner delay time for single-photon valence ionization,
and discuss the possibilities for experimental measurements.
\end{abstract}
\maketitle

\section{Introduction}

The photoelectric effect \textendash{} the emission of an electron
from matter illuminated by light - is one of the most fundamental
phenomena in nature, which historically led to Einstein's ground-breaking
proposal of the quantization of light \cite{Einstein1905} and played
a key role in the development of quantum mechanics. In the early works,
the electron emission was tacitly assumed to be instantaneous, following
the absorption of the excitation photon. However, more than half a
century ago, it was predicted theoretically that there should be a
time delay in the photoelectron emission process \cite{Wigner1955,Smith1960},
but it was only with the recent advances in attosecond science that
direct measurements of electron dynamics with attosecond time resolution
\cite{Hentschel2001} required for the experimental validation of
this prediction could be realized. Time resolved measurements of electron
dynamics were reported \cite{Uiberacker2007,Goulielmakis2010,Eckle2008,Schultze2014}
and the delay of photoemission was observed in condensed matter
\cite{Cavalieri2007} and atoms \cite{Schultze2010,Klunder2011} in the single
photon weak-field regime. However, no measurements of photoemission
time delays from molecular targets have been reported as yet. Here
we discuss theoretical results of angle and energy resolved time delays
in the photoionization of molecules, and the prospects for direct
measurement of this rich attosecond phenomena.

In scattering theory the phase of the transmitted wave is a direct
consequence of the interaction of the incident wave with the scattering
potential. Consequently, the scattering phase can be associated with
an advance or retardation of the transmitted wave caused by its interaction
with the scattering potential $V(r,\theta,\phi)$, as measured in
the asymptotic limit. This phase-shift is always relative to the $V=0$
case. A repulsive potential will lead to a negative phase, signifying
an advance of the transmitted wave, while an attractive potential
will lead to a positive phase, signifying a retardation (or trapping)
of the transmitted wave. These results are most simply derived in
a stationary state (energy-domain) picture of scattering, but a wavepacket
(time-domain) treatment yields the same essential features \cite{RTscattering}.
Hence, in a time-domain picture of photoionization, the scattering
phase-shift and associated time delay can be viewed as a group delay
of the outgoing photoelectron wavepacket, born at a time $t_{0}$
within the ionizing laser pulse. In this case, the advanced wavepacket
appears sooner than it would for the $V=0$ case, while the retarded
wavepacket appears later than it would for $V=0$. This temporal response
to the phase-shift is given by the Wigner delay, $\tau_{w}$, which
is determined by the energy-derivative of the scattering phase \cite{Wigner1955,Smith1960}.

While the concept of the Wigner delay is well established \cite{Wigner1955,Smith1960},
interest has recently been rekindled due to the experimental accessibility
of the attosecond time domain. Experiments using attosecond XUV pulse
trains or isolated attosecond XUV pulses have been able to measure the relative group delay of electron
wavepackets from atomic emission following single-photon absorption
from a weak XUV field, with the measurements additionally requiring the interaction of the electron wavepacket with an IR field  \cite{Schultze2010,Klunder2011}. The related possibility of determining an absolute photoionization time $t_{0}$ was discussed in this context \cite{Schultze2010}, and has also been explored
in the strong-field regime via tunnel ionization with ``attoclock''
measurements \cite{Eckle2008}, which employ pulses with rapidly changing
instantaneous polarization vector (e.g. circularly polarized light)
to obtain high temporal resolution via angular streaking of the photoelectron
wavepackets. 

In concert with these new experimental capabilities, numerous theoretical and computational studies have been performed. These can broadly be categorised as methodologies based on (a) canonical scattering theory \cite{Ivanov2012, Kheifets2013, Dahlstrom2013, Dahlstrom2014}, or (b) fully-numerical approaches based on the time-dependent Schrodinger equation \cite{Chacon2014, Ivanov2015}. In most cases Wigner delays from the ionization of atomic targets have been of interest, and the angle-dependence of the process has not been investigated; notable exceptions to this trend are the recent work of Wätzel et. al. \cite{Watzel2015}, who investigated the angle-dependence of the Wigner delay in detail for ionization of neon and argon, and studies of $H_{2}$ - the simplest molecular scatterer - from Serov et. al. \cite{Serov2013}, which includes some consideration of the angle-dependence \footnote{For a fuller discussion of these extant techniques and theoretical treatments, the reader is referred to recent literature from Dahlström and co-workers, in particular refs. \cite{Dahlstrom2012} and \cite{Dahlstrom2013}; also of particular relevance to this study is the recent work of Wätzel et. al. \cite{Watzel2015}, as mentioned in the main text, and related work on argon from Dahlström and Lindroth \cite{Dahlstrom2014}, which also includes discussion of the angle-dependence.}. Conceptually, these methods are of course similar - one is seeking to solve equations that determine the continuum electron wavefunctions, and obtain scattering phase-shifts. 

The main distinction that can be drawn between these approaches is the generality of the method and the information content of the results. A fully numerical treatment is, in principle, completely general, although in practice may be limited by computational cost; nonetheless, if performed carefully, the "correct" final state wavefunction should be found for any given scattering system.  A particular strength of time-dependent numerical methods is the ability to treat rapidly-varying scattering potentials, therefore allowing the effects of strong laser-fields to be incorporated into calculations. Such calculations have been employed in order to model experiments incorporating strong fields \cite{Baggesen2011,Nagele2011,Ning2014}, which cannot be treated adequately by a time-independent approach.  More traditional scattering theory approaches are usually time-independent and most suited to the weak field regime, hence are appropriate for the consideration of the intrinsic Wigner delay of the scattering system. Such approaches often use a \emph{partial-wave} formalism, which allows separation into "geometric" and "dynamical" parts. In this case much progress can be made analytically, and a deep physical insight into the characteristics of the scattering can be gained (see, for example, ref. \cite{Dill1976}). However, to obtain a complete solution to a complex scattering problem numerical methods are still ultimately needed for the dynamical part, and a specific formalism for the scattering system of interest is usually constructed in order to yield tractable equations (see, for example, refs. \cite{Ivanov2012, Serov2013}); solving molecular scattering problems is therefore non-trivial for even the simplest cases. This problem can, however, be addressed via the use of variational techniques to solve the numerical part of the problem \cite{Lucchese1986}, allowing for a methodology which retains the full physical insights of scattering theory and the generality of fully-numerical approaches, but at a significantly lower computational cost.

In this work, we investigate Wigner delays from molecular ionization based on this general approach. We explore the details of the time delay in the valence ionization
of $N_{2}$ and $CO$, based on calculations for single-photon ionization
processes. The influence of the XUV field on both the bound states
and the continuum electron are neglected, hence the results obtained
correspond to the intrinsic Wigner delays of the photoemission process
in the weak-field limit. We do not include any additional continuum-continuum
delays, which can be a significant contribution to the total observed
delay in the case of the XUV-IR measurements discussed above, but are
dependent on the experimental technique \cite{Dahlstrom2013} and
not a fundamental property of the ionizing system. In this limit,
the effect of the molecular potential on the energy and angle-resolved
Wigner delay can be explored. This fundamental exploration forms the
main thrust of the manuscript. Although the details are specific to
valence ionization of $N_{2}$ and $CO$, the results may be considered
as prototypical for molecular ionization. As detailed below (Sect. \ref{sec:calcualtions}), we make use of ePolyScat \cite{Gianturco1994,Natalense1999,Lucchese2015}, a well-developed suite of codes from the scattering community, to solve the numerical integrals for arbitrary molecular potentials, thus our methodology is completely general and can be readily applied to polyatomic
molecules. We finish by discussing some attosecond metrology concepts
which could provide deeper experimental insight into ionization time
delays in an angle and energy resolved manner.

\section{Wigner time delay}

As discussed by Wigner \cite{Wigner1955}, Smith \cite{Smith1960}
and, more recently, in some depth by various authors \cite{DeCarvalho2002,Dahlstrom2012,Pazourek2014}, the phase
of the scattering wavefunction can be associated with a time delay
of the outgoing wavepacket, $\Psi_{g}$. In a partial-wave decomposition, $\Psi_{g}$ is expressed as a coherent sume over partial-waves, $\Psi_{g}=\sum_{lm}\psi_{lm}$. Here each component is defined by the quantum numbers $(l,m)$, the electronic orbital angular momentum and its projection onto a given quantization axis respectively, and each $(l,m)$ pair defines a partial-wave scattering channel.

The time delay in a given channel is simply
the derivative of the phase with respect to energy: 

\begin{equation}
\tau_{w}(\epsilon)=\hbar\frac{d\eta_{lm}(\epsilon)}{d\epsilon}\label{eq:tauW}
\end{equation}
where $\eta_{lm}=\sigma_{l}+\delta_{lm}$ is the total scattering
phase, comprised of a central-potential (Coulombic) contribution $\sigma_{l}$
and non-central (non-Coulombic) contribution $\delta_{lm}$. For a Coulomb potential $\tau_{w}$ can be obtained
directly from $\sigma_{l}$, which can be determined analytically,
but in the general (non-Coulombic) case the total phase $\eta_{lm}$ must be determined numerically. (It is of note that this definition of the Wigner delay does not include the full $r$-dependence of the phase of the outgoing wavefunction, which is divergent for an infinite-range Coulomb potential - a more general definition incorporating the total phase is given below.  For further discussion of this point, the reader is referred to ref. \cite{Pazourek2014}, for the specific case of Wigner delays, and ref. \cite{Park1996} for a more general discussion.)

Similarly, the group delay of the outgoing electron wavepacket can
be defined as the (coherent) sum over all constituent channels:


\begin{equation}
\tau_{w}^{g}(\epsilon)=\hbar\frac{d\eta_{g}(\epsilon)}{d\epsilon}\label{eq:tauWg}
\end{equation}
Here $\eta_{g}$ represents the total (group) scattering phase, determined from $\Psi_{g}$, hence from the coherent summation over the partial-wave channels.

The significance of $\tau_{w}$ is as a time-domain manifestation
of the scattering phase $\eta_{lm}$. Both contain the same information,
namely the effect of the interaction potential on the outgoing wave,
expressed as either a phase or delay. As noted above, this definition
means that $\tau_{w}$ does not directly express the ``ionization
time'' in terms of the timescale of the interaction of the system
with a photon (or perturbing electric field), rather it describes
the time taken for the outgoing wavepacket to leave the influence
of the potential, as defined by an effective range beyond which free-particle behaviour is assumed, and expressed relative to the time taken for
a free particle with the same asymptotic velocity. In this sense a
true reference time, $t_{0}$, is only specified to be within the
duration of the ionizing radiation field.%
\footnote{While this is rigorously true for any single ionization event, it
is possible to obtain a time-resolution in practice that is better
than the pulse duration via the use of statistical sampling or, potentially,
through multi-photon ionization processes. For an ionizing pulse longer
than $\tau_{w}$ interferometric measurements which are sensitive
to the scattering phase may be used as an energy-domain proxy for
direct measurement of $\tau_{w}$. Since $\tau_{w}$ is typically
on the order of tens of attoseconds, in most cases this absolute delay
in photoemission is not measured, but rather an interferometric measurement
sensitive to the relative delay between interfering wavepackets. For
a more detailed discussion, see ref. \cite{Dahlstrom2012}.%
}

In atomic ionization, the relatively simple nature of the scattering
potential results in a continuum wavepacket with little spatial structure,
which can often be described by just two partial-wave channels. In
molecular ionization, the anisotropic nature of the potential means
that many more partial-waves are required to describe the photoelectron
wavepacket, and significant spatial and energy structure is expected.
In essence, the angular structure of the photoelectron wavepacket
is the result of the angular interferences between the partial-waves
at a given energy, while the difference in the dependence of the phase-shift
of any given $l$-wave on the photoelectron kinetic energy results
in the strong energy-dependence of the photoionization cross-section
and $\tau_{w}$.

The consequence of the angular dependence is, naturally, different
$\tau_{w}$ as a function of angle, most clearly defined in the ionizing
or molecular frame. We can rewrite equation \ref{eq:tauWg} for this
more general case:

\begin{equation}
\tau_{w}(k,\theta,\phi)=\hbar\frac{d\arg(\psi_{lm}^{*}(k,\theta,\phi))}{d\epsilon}\label{eq:tauW_mol}
\end{equation}
In this case we explicitly write $\tau_{w}$ as a function of the
partial-waves $\psi_{lm}(k,\theta,\phi)$, labelled as a function
of photoelectron momentum $k$, and polar and azimuthal angles $(\theta,\phi)$
relative to the molecular axis. These wavefunctions contain both the
scattering phase $\eta_{lm}(k)$ plus an angular contribution $Y_{lm}(\theta,\phi)$.
The complex conjugate is required here because the scattering phase
appears as $e^{-i\eta_{lm}}$ in $\psi_{lm}$ (for a discussion of
continuum wavefunctions in photoionzation, see ref. \cite{Park1996}).
As before, this equation expresses $\tau_{w}$ for each partial wave
channel, and the group delay results from the sum over all $(l,m)$
terms:

\begin{equation}
\tau_{w}^{g}(k,\theta,\phi)=\hbar\frac{d\arg(\sum_{l,m}\psi_{lm}^{*}(k,\theta,\phi))}{d\epsilon}\label{eq:tauW_mol_sum}
\end{equation}

In this work we examine the form of the energy and angle-resolved
group delay for two specific benchmark cases, valence ionization of
the diatomic molecules $N_{2}$ and $CO$, and consider how the delay
responds to the details of the molecular potential and the resulting
continuum wavefunction.

\section{Group delay in the molecular frame \label{sec:calcualtions}}
\begin{figure}
\includegraphics[width=0.95\columnwidth]{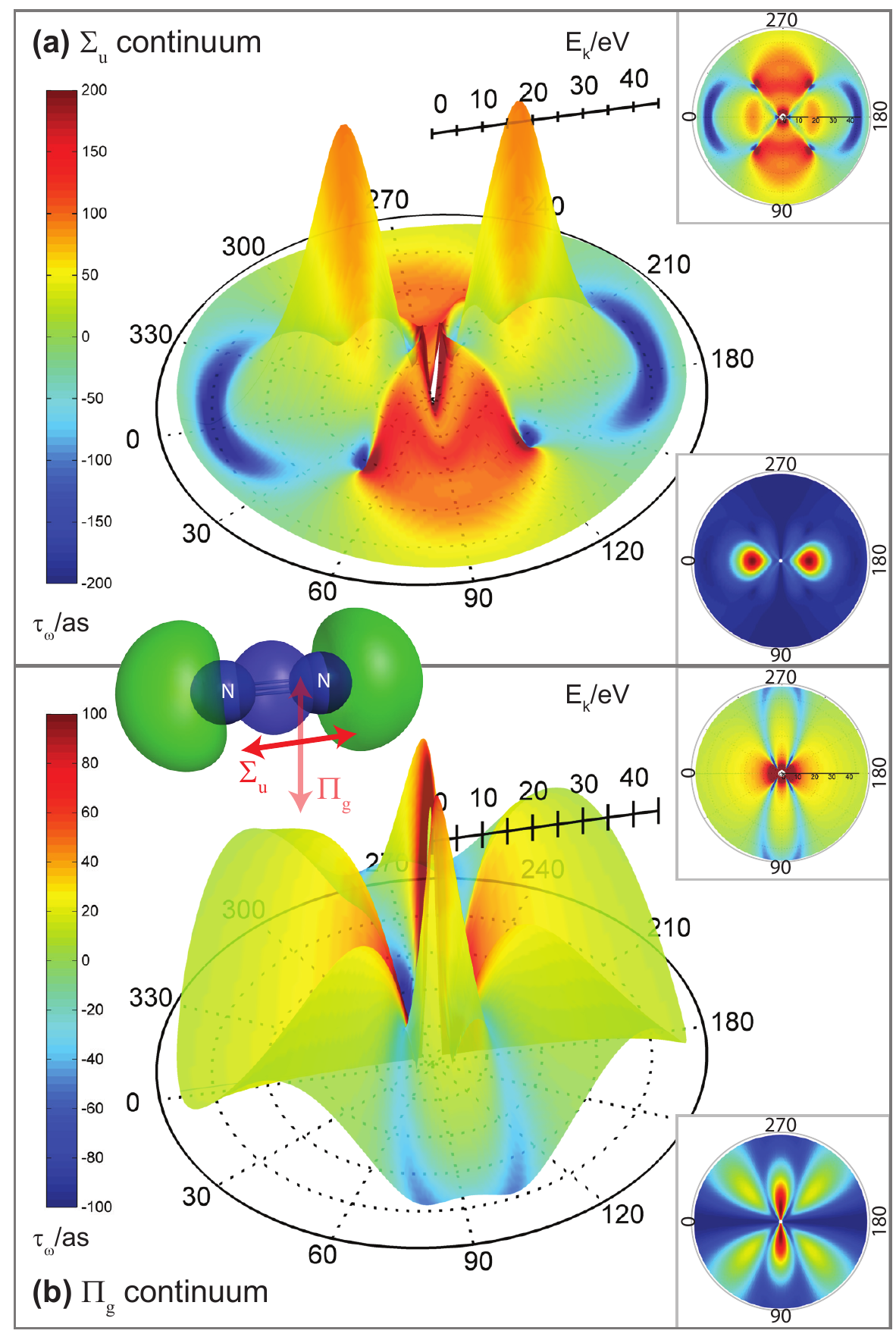}

\caption{Group delays for ionization of $N_{2}$ ($3\sigma_{g}\rightarrow k\sigma_{u},\, k\pi_{g}$).
(a) $\Sigma_{u}$ continuum, (b) $\Pi_{g}$ continuum. The main plots
show polar surfaces, as a function of photoelectron kinetic energy
and angle, with the topography defined by the photoionization cross
section and colour-map by the Wigner delay $\tau_{w}^{g}$; insets
show the same data as 2D polar colour-maps, upper plot for $\tau_{w}^{g}$
(same scale as main colour-map) and lower plot for photoionization
cross-sections (arb. units).\label{fig:Group-delays-for-N2}}

\end{figure}

\begin{figure}
\includegraphics[width=0.95\columnwidth]{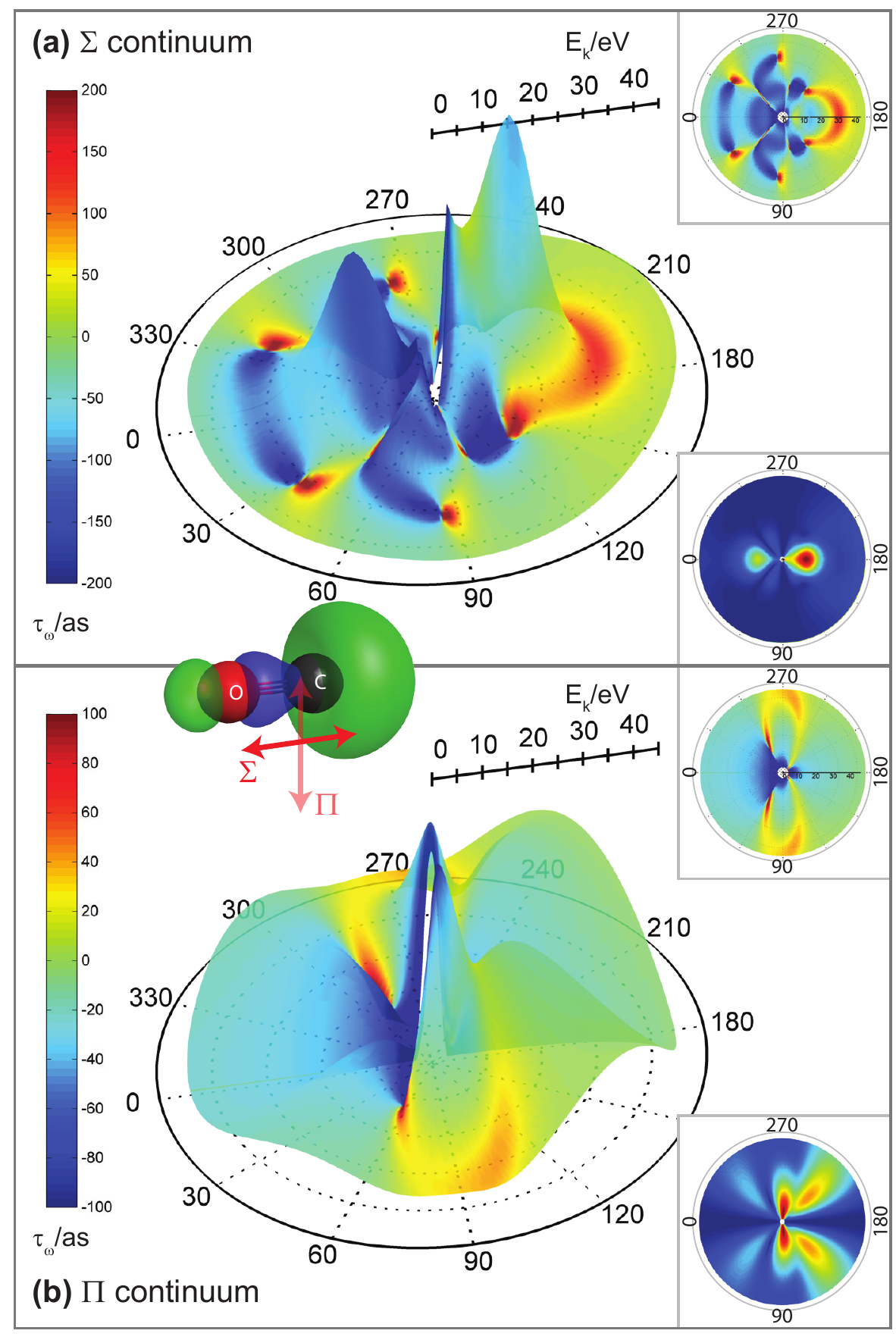}

\caption{Group delays for $CO$ ($5\sigma\rightarrow k\sigma,\, k\pi$). (a)
$\Sigma$ continuum, (b) $\Pi$ continuum. The main plots show polar
surfaces, as a function of photoelectron kinetic energy and angle,
with the topography defined by the photoionization cross section and
colour-map by the Wigner delay $\tau_{w}^{g}$; insets show the same
data as 2D polar colour-maps, upper plot for $\tau_{w}^{g}$ (same
scale as main colour-map) and lower plot for photoionization cross-sections
(arb. units).\label{fig:Group-delays-for-CO}}

\end{figure}

\subsection{Numerical details}
Ionization matrix elements, which include the full scattering phase,
were calculated using the ePolyScat suite of codes, distributed by
R. R. Lucchese (for further details see refs. \cite{Gianturco1994,Natalense1999,Lucchese2015}).
These calculations take input from standard electronic structure codes
(Gamess, Gaussian etc.) to define the initial state of the system.
Ionization is treated as a 1-electron process, leading to an $N-1$
electron system and a free electron (hence there are no multi-electron
effects in the sense of core relaxation, polarization etc.). The continuum
wavefunction is solved numerically in the $N-1$ electron potential,
via a Schwinger variational procedure \cite{Lucchese1986}, and ionization
matrix elements (within the dipole approximation) are calculated as
the spatial overlap of this wavefunction and the initial orbital wavefunction,
for a given polarization of the light and at single photoelectron
energy. This approach has been shown to work well in the weak field
regime \cite{Lucchese1986}, and also for calculation of recombination
matrix elements in HHG \cite{Le2009} although, in general, it is
not an appropriate technique for the strong field regime as the laser
field is not included in the scattering calculations.

In this work, calculations were based on equilibrium geometries and
electronic structure from Gamess calculations (run at a relatively
low, but appropriate, level of theory: RHF/MP2/6-311G) \cite{gamess},
with equilibrium bond lengths found to be 1.07~Å ($N_{2}$) and 1.12~Å
($CO$). Continuum wavefunctions and dipole matrix elements were computed
with ePolyScat, for the highest-lying $\sigma$-orbitals in both cases,
for linearly polarized ionizing radiation in both parallel and perpendicular
geometries, and for photoelectron energies from 1 to 45~eV.  The phase
information from the raw matrix elements, expressed in terms of angular
momentum channels, provides the full scattering phase-shift, and application
of eqn. \ref{eq:tauW_mol} provides $\tau_{w}$ for each channel.
Similarly, eqn. \ref{eq:tauW_mol_sum} provides the group, or photoelectron
wavepacket, delay. In the calculations, radial integrals are evaluated for $r_{max}=$~10~Å, defining an effective range to the interaction at which the total phase (hence delay) is defined. By calculating the photoionization matrix elements
for a range of photoelectron energies, the energy dependence of the
process can be mapped out, and the complete dependence of the Wigner
delay $\tau_{w}(k,\theta,\phi)$ obtained. 

In the following, we present and discuss these results for the general reader. Supplementary materials, including additional technical details of the results, e.g. channel-resolved dipole matrix elements, which may be of interest to some readers, are available online via Figshare at \url{http://dx.doi.org/10.6084/m9.figshare.2007486}.

\subsection{Results}
The results for the group (channel-integrated) Wigner delay, $\tau_{w}^{g}(k,\theta,\phi)$,
for nitrogen and carbon monoxide are shown in figures \ref{fig:Group-delays-for-N2}
and \ref{fig:Group-delays-for-CO}, and represent the main results
of this work. In the standard notation of \emph{ionizing orbital$\rightarrow$continuum
wave} the 1-electron ionization channels are given as $3\sigma_{g}\rightarrow k\sigma_{u},\, k\pi_{g}$
for $N_{2}$ and $5\sigma\rightarrow k\sigma,\, k\pi$ for $CO$ \cite{Lucchese1986}.
In the following discussion these cases are denoted by the overall
$N$-electron symmetry ($\Gamma_{ion}\bigotimes\Gamma_{electron}$)
and species, e.g. $N_{2}(\Sigma_{u})$, $CO(\Pi)$ etc.

Due to the cylindrically symmetric nature of these molecules,
the $\phi$-coordinate is redundant in these cases, and we can show
the complete results as polar surface plots, as a function of energy
and angle $\theta$ (relative to the molecular axis), without any
loss of information. In these plots the surface topography follows
the magnitude of the dipole matrix element (proportional to the square-root
of the photoionization cross-section), while the colour-map shows
the energy and angle-resolved Wigner time. As an alternative presentation
of the results, which may be clearer in print, the insets show the
same data as polar colour-maps. The $\Sigma$ and $\Pi$ continua
shown correspond to parallel or perpendicular laser polarization in
the molecular frame respectively. The difference in peak magnitude
between the continua is not shown in the figures, which are independently
normalised to emphasize the angular structure, but it is of note that
the $\Sigma$ continua dominate in both cases, with the peak magnitude
ratios of $\sim$2.4:1 for $N_{2}(\Sigma_{u}):N_{2}(\Pi_{g})$ , and
$\sim$5.3:1 for $CO(\Sigma):CO(\Pi)$. The molecular structure and
ionizing orbital are also shown for reference, and the laser polarization
correlated with the different photoionization continua accessed are
indicated.

These results present a complete, but complicated, picture of the
molecular photoionization event, and the associated Wigner delay for
the outgoing photoelectron wavepacket. It is immediately apparent
that there is a significant amount of structure observed, both as
a function of energy and angle, with $\tau_{w}^{g}$ values ranging
from -200 to +200~as.

In both cases, the ionizing orbital is the valence $\sigma$-bonding
orbital, with lobes oriented along the molecular axis. The choice
of polarization of the ionizing radiation - either parallel or perpendicular
to the molecular axis - defines the symmetry of the ionization continuum
accessed, hence the symmetry of the continuum photoelectron wavefunction.
For both $N_{2}$ and $CO$, this results in peaks in the cross-section
along the molecular axis ($\theta=0^{\circ},\,180^{\circ})$ for the
parallel case (figs. \ref{fig:Group-delays-for-N2}(a) and \ref{fig:Group-delays-for-CO}(a)),
and orthogonal to the molecular axis ($\theta=90^{\circ},\,270^{\circ})$
for the perpendicular case (figs. \ref{fig:Group-delays-for-N2}(b)
and \ref{fig:Group-delays-for-CO}(b)). Weaker additional lobes are
also observed in all cases, but are most pronounced in the $CO(\Pi)$
case, where they peak only around 20~\% lower than the perpendicular
features. Furthermore, the lack of inversion symmetry in $CO$ results
in a significant difference in the cross-sections between the oxygen
($\theta=0^{\circ})$ and carbon ($\theta=180^{\circ})$ ends of the
molecule, which is clear in both the $\Sigma$ and $\Pi$ continua,
and again particularly pronounced in the additional lobes in the $\Pi$
case, which dominate the cross-section around the carbon end of the
molecule ($\theta=140^{\circ},\,230^{\circ})$.

\section{Scattering dynamics}

Physically, the peaks in the cross-section correspond to maxima in
the dipole integrals which define the coupling between initial orbital
and final continuum wavefunctions induced by ionizing radiation, with
an angular dependence given by the partial-wave interferences. For
$N_{2}(\Sigma_{u})$ this peak is the well-known shape-resonance \cite{Lucchese1986,Shigemasa1995,Hikosaka2000},
corresponding to an enhancement of the $l=3$ partial-wave, which
can be considered as a trapping of this part of the outgoing wavepacket
due to the form of the molecular potential energy surface. It is therefore
not unexpected that the Wigner delay is also long in this region.
Less expected are the lobes almost perpendicular to the molecular
axis seen in fig. \ref{fig:Group-delays-for-N2}(a), and associated
long delays. This can be physically rationalized as a trapping of
the outgoing wave in the bonding region (i.e. the nitrogen-nitrogen
triple bond), resulting in  a long Wigner delay. For $N_{2}(\Pi_{g})$
the symmetry of the problem results in a nodal plane along the molecular
axis, so there is much reduced overlap between the main lobes of the
ionizing orbital and the $\Pi_{g}$ continuum, as compared to the
$\Sigma_{u}$ case. Here the cross-section looks akin to scattering
through a slit, with a main feature and lower-intensity side lobes,
and the cross-section peak is significantly reduced as compared to
the $\Sigma_{u}$ case, as discussed above. The dipole integral peaks
much closer to 0~eV, and it is only in the low-energy region that
large Wigner delays are predicted. For most of the energy and angular
range the Wigner delay is close to zero, consistent with a classical
diffractive picture of the ionization event, in which there is little
trapping of the outgoing photoelectron wave.

In the case of $CO$ the picture is quite different. Here the Wigner
delays are predominantly negative, indicating a slight net repulsive
effect from the molecular potential, and the results are highly asymmetric,
consistent with the loss of inversion symmetry and the form of the
ionizing orbital for a polar diatomic. The repulsive nature of the
potential is most significant at the oxygen end of the molecule, where
the extent of the ionizing orbital is much reduced relative to the
carbon end. Chemically, the small extent of the orbital signifies
the ``electronegativity'' of the oxygen atom, which will tend to
acquire a slight negative charge relative to the carbon atom. Based
on chemical intuition, one might therefore expect to find a more repulsive
potential than for the carbon end of the molecule, and this is borne
out in the Wigner delay results. At higher energies, the Wigner delay
at the carbon end becomes positive and large. This can be understood
by consideration of the radial part of the continuum wavefunction:
at higher energies the photoelectron wavelength becomes shorter, and
the continuum function will become more penetrating relative to the
core wavefunction. Consequently, the spatial overlap integral will
incorporate more bound-state density closer to the core, which is
effectively more strongly bound due to the slightly positive overall
charge over the carbon atom, and will thus be delayed relative to
bound-state density far from the core. At the oxygen end, the same
change in overlap has the opposite effect, and continues to result
in large negative Wigner delays due to the repulsive nature of the
molecular potential over a large spatial region. 

\begin{figure}
\includegraphics[width=0.95\columnwidth]{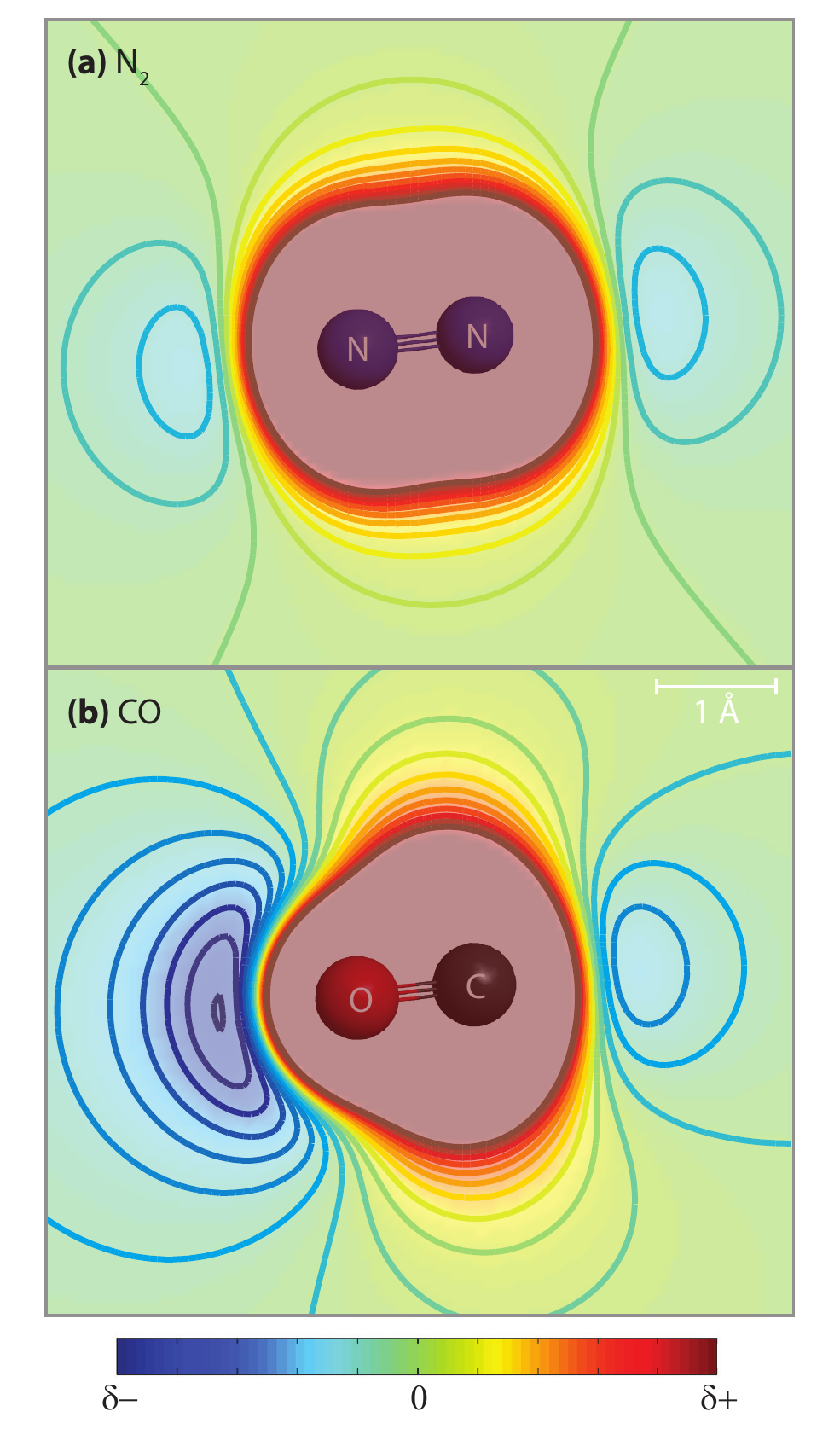}

\caption{Molecular electrostatic potentials. (a) $N_{2}$, (b) $CO$. Contours
show the long-range part of the molecular potential, with the colour
scale indicating slightly positive ($\delta+$) and slightly negative
($\delta-$) regions.\label{fig:Molecular-electrostatic-pots}}

\end{figure}

In order to visualize this behaviour, figure \ref{fig:Molecular-electrostatic-pots}
shows the molecular electrostatic potentials $V(r,\theta)$ for both (neutral)
molecules \cite{Bode1998,Bode}. In the figure, a cut through the
cylindrically symmetric potentials are shown by both a colour-map
and contours. The ranges plotted are chosen to highlight the long-range
part of the potential which is most structured, and largely responsible
for the complexity of the scattering problem. The short-range, highly
positive, part of the potential, within which the majority of the
bound electronic population resides, therefore appears structureless
in these figures. Here it is clear that the negative, repulsive part
of the potential is much more significant for $CO$ than for $N_{2}$,
and most significant around the oxygen end of the molecule, thus leading
to the most pronounced negative Wigner delays for wavepackets which
experience this region. Conversely, the primarily attractive or neutral
nature of the scattering potential for $N_{2}$, is responsible for
the positive Wigner delays observed in the calculations.

\begin{figure}
\includegraphics[scale=1.1]{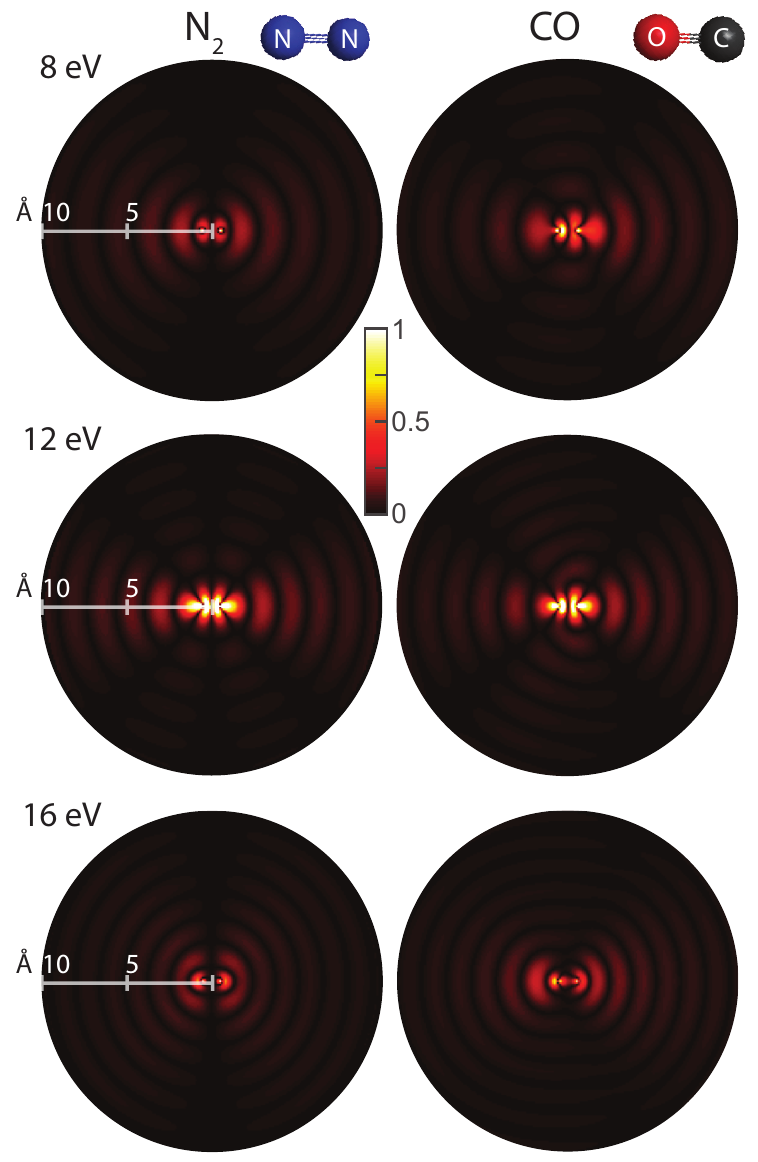}

\caption{Continuum wavefunctions $|\psi(\theta,\phi)|$ for scattering from
$N_{2}(\Sigma_{u})$ and $CO(\Sigma)$ at $E=8,\,12,\,16$~eV. Each
plot is normalized to the peak of the wavefunction to highlight the
spatial structure. \label{fig:Continuum-wavefunctions}}

\end{figure}

Visualization of the scattering wavefunctions provides additional
physical insight into the dynamics of the process. Figure \ref{fig:Continuum-wavefunctions}
shows a selection of continuum wavefunctions at different photoelectron
energies, chosen to represent the evolution of the scattering wavefunctions towards the peak in the cross-sections (shape-resonance), with symmetries concomitant with ionization parallel to
the molecular frame ($N_{2}(\Sigma_{u})$ and $CO(\Sigma)$).  At
the highest energy shown, the far-field wave-front (approximately
established at length-scales as short as several Å \cite{Knox2010})
shows little obvious angular structure correlated with the core, save
for a basic 2-centre scattering pattern. In contrast, at the two lower energies
the angular structure is more complex, with the nodal planes more
pronounced. This change in angular structure, for $N_{2}$, is exactly
the shape-resonance effect discussed above, with the observed continuum
structure corresponding to the rise and fall of the $l=3$ partial-wave
component over this energy range, including a significant change in
the magnitude of the wavefunction in the core region which has a strong
effect on the overall ionization yield. For $CO$ the effect is slightly
less clear, since the continuum structure is more complicated, but
the general trend in complexity of the angular structure with energy
is similar, and has been labelled by other authors as a shape resonance
analogous to the $N_{2}$ case \cite{Lucchese1986}. In all cases,
the asymptotic phase-shift of the waves is approximately established
at the length-scales shown ($r_{max}=$10~Å), and phase differences
can be observed in the plots. The lack of inversion symmetry in the
far-field phases for $CO$ is clear, with the phase shift between
the carbon and oxygen ends of the molecule apparent in the intensity
at the 10~Å cut-off.

Most generally, the complex structures observed for these two, relatively
simple, diatomics might be regarded as indicative of molecular photoionization
from valence orbitals, which invariably involves spatially diffuse,
highly structured wavefunctions. The nature of the molecular potential,
which is responsible for the shape of the bound-state orbitals, will
similarly result in a continuum scattering wave which is highly sensitive
to angle and energy. In this particular set of results, the effect
of symmetry-breaking along the molecular axis is very clear, and in
general larger molecules with lower symmetry may be expected to show
similar, asymmetric, highly structured photoionization delays. The
angular-sensitivity of the results points to the importance of angle-resolved
measurement (or, equivalently, the loss of information inherent in
angle-integrated measurements) for the investigation of molecular
photoionization delays, and we consider this further in the following
section.

\section{Measurement\label{sub:Measurement}}

Recent measurements in atomic ionization using attosecond pulses have
shown how $\tau_{w}$ can be measured in the time-domain. Ionization
with attosecond XUV pulses, probed via streaking measurements \cite{Schultze2010},
and side-band measurements \cite{Klunder2011} have been demonstrated.
In both cases the effect of the IR probe field on the measurement
is significant, and its effect on the photoelectron must be taken
into account in order to model (or extract) and understand the measured
delays. A series of theory papers have also discussed this issue (for
example refs. \cite{Kheifets2010,Baggesen2011,Nagele2011,Dahlstrom2013,Ning2014,Chacon2014}),
most recently considering the angle-dependence of the time-delays
in atomic ionization \cite{Dahlstrom2014,Watzel2015}, and left-right asymmetry in molecular ionization for $CO$ \cite{Chacon2014}.

In essence, the measurements work by mixing the electron wavepackets
with the IR field, creating a spectrogram with modulations referenced
to the carrier-envelope phase of the IR pulse. The streaking experiments
used a strong IR field and FROG reconstruction of the resulting spectrogram
in order to determine the delay between photoelectrons emitted from
different initial states (and at different energies). The side-band
measurement is based on single-photon absorption or emission in order
to interfere photoelectrons from the same initial state, but created
at different energies via different harmonic orders in the pump pulse
train. This is effectively the RABBIT technique \cite{Muller2002},
but implemented to obtain photoelectron scattering phases instead
of optical phase information as per its original conception %
\footnote{Interestingly, ref. \cite{Muller2002} notes that the electron scattering
phase contribution to the RABBIT measurement can be ``easily taken
into account'' since it can be ``calculated from atomic theory with
very good accuracy''. In the context of atomic ionization of simple
species ($H$, $He$) this is reasonable, but does not hold for many
electron systems, and certainly not for molecular scatterers.%
}. For example, ref. \cite{Haessler2009} investigated the effect of ionization resonances on the phases obtain from RABBIT studies of molecular nitrogen. In essence, these types of measurement rely on phase differences
between the interfering photoelectron wavepackets, so are sensitive
to the difference in the group delays between different photoelectron
energies, however they are angle-averaged over the photoelectron emission
direction in the lab frame, and all partial-wave components. 
It is of particular note that the angle-resolved cross-section will weight the angle-integrated measurement towards the Wigner delays of the main angular features. \footnote{Angle-integrated Wigner delays corresponding to the exemplar cases presented herein can be found in the online Supplementary Materials at \url{http://dx.doi.org/10.6084/m9.figshare.2007486}.}

The scattering phases of individual partial waves, at a single energy,
can be determined by measurements of photoelectron angular distributions.
These are usually termed ``complete'' photoionization experiments,
and have been successfully demonstrated for a range of atomic and
molecular ionization process (see refs. \cite{Duncanson1976,Reid1992}
for example, for more comprehensive reviews see refs. \cite{Reid2003,Reid2012}),
and most recently for multi-photon ionization with femto-second pulses,
including electronic dynamics \cite{Hockett2014}. However, these
measurements are typically not able to ascertain the phase structure
with respect to energy, so can only determine $\eta_{lm}$ for a given
set of partial-waves with one of the waves serving as a reference.
These types of measurement therefore provide detailed information
on the angular part of the problem, including the phases of the contributing
partial-waves, but do not directly provide a full mapping of $\tau_{w}(k,\theta,\phi)$
\footnote{Although this is strictly correct, it is the case that careful analysis
of PADs recorded at different energies, possibly combined with guidance
from theory, can provide phase information as a function of energy.
See, for example, ref. \cite{Yagishita2005} . With such an approach
the full energy and angle-dependent $\tau_{w}(k,\theta,\phi)$ could
be obtained.%
}. The possibility of such experiments in the atto-second regime has
also yet to be explored, although it is feasible that the broad energy
bandwidths available would allow for phase structure as a function
of energy to also be determined.

Ultimately, a combination of these techniques would be capable of
measurements of the full $\tau_{w}^{g}(k,\theta,\phi)$. An angle-resolved
RABBIT methodology would provide the energy and angular dependence
of $\tau_{w}^{g}$ %
\footnote{Interestingly, this proposition was already suggested in ref. \cite{Muller2002},
but for the purpose of providing background free side-bands.%
}, and measurements in this framework have very recently been investigated
for atomic ionization \cite{Laurent2014,Heuser2015,Heuser2015a,Ivanov2015}.
A detailed analysis of the photoelectron angular distributions - possibly
from the same measurements, or more simply via direct ionization measurements
- could provide complementary partial-wave information; the coupling
of these two analyses could thus provide $\tau_{w}(k,\theta,\phi)$.
The cleanest measurement strategy would also make use of molecular
alignment, in order to choose only a single continuum symmetry. This
would increase the complexity of the experiment, but allow for a decrease
in the complexity of the analysis.

Very similar considerations have been explored in the context of high-harmonic
generation (HHG). In particular, angle and energy resolved phase measurements
of $Br_{2}$ were performed with the LAPIN technique \cite{Bertrand2013}.
In this technique, a two part measurement strategy (similar to that
outlined above) is used in order to provide data which allows for
reconstruction of the energy and angular dependence of the phase of
the emitted high-harmonic radiation. In this case the measured emission
phase includes contributions from strong-field ionization, propagation
in the continuum and photo-recombination (this is the three-step model
of HHG); the final step here is effectively equivalent to single-photon
ionization. The experiments are based on two-source interferometry
techniques: sensitivity to the angle-dependence of the phase is obtained
in the case of two spatially distinct harmonic sources, both with
\textbf{$Br_{2}$ }molecules, but with one source aligned and the
other unaligned; sensitivity to the energy dependence of the phase
is obtained in the case of two distinct species of emitter, with harmonics
generated from a mixed gas containing $Br_{2}$ and a reference atom
($Xe$). The combination of the measurements, combined with a self-consistent
phase-reconstruction procedure, provided angle and energy-dependent
phase information. The reconstructed phases agreed reasonably well
with theoretical results, which were based on ePolyScat calculations
similar to those employed herein. Although a relatively involved procedure,
the complete phase information obtained with the LAPIN technique will
contain $\tau_{w}^{g}(k,\theta)$ from the recombination process,
however other sources of delay will be present.

Another related study from the field of HHG is that of measurements
on oriented $CO$, which was combined with a theoretical treatment
(again within the standard three-step model) in order to understand
the various phase contributions to the emitted harmonics \cite{Frumker2012}.
In this case the prediction of even harmonics relied on both the difference
in phase between the ends of the molecule, and the phase accrued during
the tunnel ionization and propagation steps (also directionally dependent
due to the shape of the molecular potential). Although these measurements
are made in the frequency domain, the process can be understood in
the time domain as attosecond bursts of harmonics occurring on each
half-cycle of the driving laser field. The spectral interference of
these bursts at the detector (hence integrated over the driving pulse
duration and the generation volume) then determines the magnitude
of the harmonics. Although this mechanism is responsible for all harmonic
generation, in the CO experiments it is especially pertinent for understanding
the effect of the asymmetry of the molecular potential, which results
in different timings of the ionization and recollision leading to
a phase difference which is mapped to the generated XUV bursts. In
such measurements the global phase structure can be ascertained due
to the bandwidth, or energy multiplexing, present. Although the spectral
phase in this case was not measured directly, calculations based on
a modified 3-step model using time-dependent ionization and propagation
calculations, combined with accurate recombination matrix elements
(hence scattering phases) were able to recreate the intensity envelope
of the harmonic spectrum and spectral phase differences between opposites
end of the molecule. While not providing the full mapping of $\tau_{w}^{g}(k,\theta,\phi)$
discussed above, these types of measurements are very sensitive to
phase differences in specific directions ($\theta=0^{\circ}$ vs.
$\theta=180^{\circ}$ in this case), and could therefore provide an
interesting step towards full angle-resolved time-delay measurements,
with the benefit of significantly reduced experimental complexity.
One might consider that this frequency domain measurement of coherent
attosecond processes is a technique sensitive to dynamics on the time-scale
of $\tau_{w}$, so could additionally be a sensitive probe of electronic
dynamics.

A final point of note with regard to measurement of molecular vs.
atomic Wigner delays is the increased density of states in the molecular
case. This suggests that the main difficulty in application of the
measurement schemes discussed above, particularly to polyatomics,
will likely be spectral congestion due to overlapping vibronic bands
in the photoelectron spectrum. There is no general solution to this
problem, since it is somewhat inherent to molecular ionization, but
in many cases the issue may be side-stepped by judicious choice of
spectral window(s) for RABBIT or similar type measurement schemes,
with the obvious cost of reducing the energy range which can be investigated.
Alternatively, it may be possible to make use of this additional structure
by, for instance, probing the effects on the Wigner delay of ionizing
from different states, or via different intermediate states by making
use of degenerate ionization processes of different photon orders.
In both cases, degenerate photoelectrons will interfere (providing ensemble coherence is maintained, and the process is symmetry-allowed), and information
on the Wigner delays associated with the different ionizing transitions
will be contained in the measurement. Conceptually this is similar
to the measurements of ref. \cite{Schultze2010}, which ascertained
the difference in Wigner delay between photoelectron wavepackets originating
from $2s$ and $2p$ ionizing states. However, in that case the measurement
was made via streaking of energetically separated photoelectron bands,
rather than via direct interference between the bands. Older frequency-domain
work has investigated exactly the case of degenerate photoelectron
band interferences suggested here, examples include coherent control
and complete experiments \cite{Yin1992,Wang2001}, and as a method
sensitive to the Breit-Wigner phase shift of an intermediate bound-state.
However, in these cases only narrow energy ranges were considered,
so these older works did not consider the energy-dependence of the
photoionization phase and the associated Wigner delays.

\section{Conclusions}

Molecular ionization is a complex phenomenon, with the outgoing photoelectron
wavepacket experiencing a highly anisotropic scattering potential.
In the time-domain, this results in a highly-structured Wigner delay,
as a function of energy and angle in the molecular frame. With the
use of scattering calculations, the angle-dependent Wigner delay $\tau_{w}^{g}(k,\,\theta,\,\phi)$
was examined for two simple diatomics, and these results illustrate
the magnitudes of the delays, and types of structures, which might
generally be expected in molecular photoionization. The deep link
between the Wigner delay and the photoionization matrix elements is
also revealed in the correlation of energy-domain photoionization
phenomena - in this case the shape resonance in $N_{2}$ - with features
in the Wigner delay. Physically, this correspondence arises from
the mildly attractive and repulsive regions in the long-range part
of the scattering potential, which largely determine the continuum
photoelectron wavefunction at the energy ranges investigated. In a
wavepacket picture, the same considerations are manifested as large
changes in the photoelectron wavepacket dwell-times in these spatial
regions, both as a function of energy and angle in the molecular frame.
Finally, some concepts for the experimental measurement of angle-resolved
Wigner delays were discussed, suggesting the possibility of experimental
methodologies based on existing RABBIT measurements (and conceptually
similar HHG studies) for the measurement of angle-resolved Wigner
delays. While the outlook here is promising, given the highly-structured
nature of the Wigner delay and molecular ionization continuum, such
experiments will be very challenging.

\section*{Acknowledgements}

EF acknowledges support by the Israel Science Foundation (grant No.
1116/14) and European Commission Marie Curie Career Integration Grant.
PBC acknowledges support from Canada's NSERC, the US Air Force Office
of Scientific Research as well as the from Multidisciplinary University
Research Initiatives from the US Army Research Office (WN911NF-14-1-0383).

\bibliographystyle{unsrt}
\bibliography{ionization_time_delay,ionization_time_delay_210715,gamess,additional_refs_300915,additional_refs_080216}

\section*{}
\end{document}